\documentclass[fleqn,usenatbib]{mnras}

\usepackage{newtxtext,newtxmath}

\usepackage[T1]{fontenc}
\usepackage{ae,aecompl}

\usepackage{graphicx}	
\usepackage{amsmath}	
\usepackage{bm}
\usepackage{multirow}
\usepackage{graphicx}
\usepackage{stfloats}

\newcommand{\shmr}{$M_*$--$M_{\text{halo}}$}

\title[Satellite Luminosity Functions]{The Luminosity Functions and Redshift Evolution of Satellites of Low-Mass Galaxies in the COSMOS Survey} 

\author[Roberts et al.]{
Daniella M. Roberts,$^{1,2}$\thanks{E-mail: roberts.1611@osu.edu}
Anna M. Nierenberg,$^{4}$
Annika H. G. Peter $^{1,2,3}$
\\
$^{1}$Department of Physics, The Ohio State University, 191 W. Woodruff Ave., Columbus, OH 43210, USA \\
$^{2}$Center for Cosmology and AstroParticle Physics, The Ohio State University, 191 W. Woodruff Ave., Columbus, OH 43210, USA\\
$^{3}$Department of Astronomy, The Ohio State University, 140 W. 18th Ave., Columbus OH 43210, USA\\
$^{4}$Department of Physics, University of California Merced, 5200 North Lake Rd. Merced, CA 95343\\
}

\date{Accepted XXX. Received YYY; in original form ZZZ}

\pubyear{2020}

\begin{document}
\label{firstpage}
\pagerange{\pageref{firstpage}--\pageref{lastpage}}
\maketitle

\begin{abstract}

The satellite populations of the Milky Way, and Milky-Way-mass galaxies in the local universe, 
have been extensively studied to constrain dark-matter and galaxy-evolution physics. 
Recently, there has been a shift to studying satellites of hosts with stellar masses between that of the Large Magellanic Cloud and the Milky Way, since they can provide further insight on hierarchical structure formation, environmental effects on satellites, and
the nature of dark-matter.
Most work is focused on the Local Volume, and little is still known about low-mass host galaxies at higher redshift.
To improve our understanding of the evolution of satellite populations of low-mass hosts, we study satellite galaxy populations as a function of host stellar mass \mbox{$9.5<\log(M_*/M_{\odot})<10.5$} and redshifts \mbox{$0.1 < z < 0.8$} in the COSMOS survey, making this the first study of satellite systems of low-mass hosts across half the age of the universe. 
We find that the satellite populations of low-mass host galaxies, 
which we measure down to satellite masses equivalent to 
the Fornax dwarf spheroidal satellite of the Milky Way, remain mostly unchanged through time. We observe a weak dependence between host stellar mass and number of satellites per host, which suggests that the stellar masses of the hosts
are in the power-law regime of the stellar mass to halo mass relation $(M_*-M_{\text{halo}})$ for low-mass galaxies. 
Finally, we test the constraining power of our measured cumulative luminosity function to calculate the low-mass end slope of the $M_*-M_{\text{halo}}$ relation.
These new satellite luminosity function measurements are consistent with $\Lambda$CDM predictions.

\end{abstract}




\begin{table*}
\caption{The number of host galaxies in each of the four data bins.}
\label{tab:host_sample}
\hskip-1.0cm
\begin{tabular}{cccc}
                                                                              & \multicolumn{1}{l}{} & \multicolumn{2}{c}{Redshift Range} 
                                                                              \vspace{0.1cm}
                                                                              \\
                                                                              \cline{3-4}
                                                                             
                                                                              \vspace{0.1cm}
                                                                              &                      & $0.1 < z < 0.4$  & $0.4 < z < 0.8$ 
                                                                              \\ \cline{2-4} 
                                                                              \vspace{0.1cm}
\multirow{2}{*}{\begin{tabular}[c]{@{}c@{}}Stellar Mass\\ Range\end{tabular}} & $  9.5 < \log ( M_*/M_{\odot}) < 10.0 $   & 736              & 1539            \\ 
& $ 10.0 < \log (M_*/M_{\odot}) < 10.5 $  & 614              & 1375            \\ \cline{2-4} 
\end{tabular}%
\end{table*}

\section{Introduction}
\label{sec:intro}

The cosmology that underlies our models of galaxy formation is dominated by dark energy, represented by the cosmological constant $\Lambda$, cold dark matter (CDM), and includes a small baryonic matter component \citep{blumenthal1984formation, ade2016planck}. 
In this model, the early universe produced small density perturbations with amplitudes that became increasingly amplified at decreasing length scales. \citep{zeldovich1970,Peebles1980,Guth:1980zm,Linde:1981mu,Albrecht:1982wi}. Thus, the smallest perturbations were the first to collapse and evolve through hierarchical merging to form large structures, such as virialized dark matter halos. 
The small baryonic component in $\Lambda$CDM also merged hierarchically within these halos, forming the visible galaxies we see today \citep{springel2006large}. 

Galaxies within their own dark matter halos can become
gravitationally bound to a more massive host galaxy. This merging strips away the smaller halo's outer region, while leaving its central region intact for much longer. This allows smaller galaxies to survive within the virialized regions of the larger halo \citep{berezinsky2008remnants}.   We observe these accreted galaxies as satellite galaxies residing within their own halo substructure, called subhalos \citep{davis1985,natarajan1998mass, diemand2007formation}.

Because halos and galaxies form hierarchically, we expect satellite galaxies to be a ubiquitous feature of central galaxies. 
These satellite galaxies are mapped to their corresponding dark matter halos through
the stellar to halo mass relation \shmr; which is predicted to have a power-law dependence $M_* \propto M_{\text{halo}}^{\beta}$ approaching dwarf-galaxy scales $M_* \lesssim 10^{10} M_{\odot}$ \citep{guo2010galaxies,behroozi2013average,behroozi2019,moster2010constraints,moster2013,moster2018,munshi2013,munshi2019,read2017stellar,rodriguezpuebla2017,mowla2019mass,wheeler2019}. The shape of this relation is expected to be established at high redshift.
Since dark-matter substructures have a scale-free mass function \mbox{$dN_{\text{subhalo}}/dM_{\text{subhalo}}\propto M_{\text{subhalo}}^{\alpha}$} at the low-mass end, then together with the amplitude and shape of the satellite stellar mass function, or luminosity function \mbox{$dN_{\text{satellite}}/dL_{\text{satellite}}$}, one can find the slope $\beta$ of the \shmr~relation. Therefore, satellite galaxy abundances can be related to a host galaxy's halo mass \citep{Sales2013}, making observations of satellite galaxies---as a function of host mass, luminosity, and time---important tests of the hierarchical structure formation paradigm of $\Lambda$CDM and galaxy evolution.

This model is successful at describing large scale structures \mbox{($\gtrsim 10$ Mpc)} in the universe \citep{cole20052df, komatsu2011seven, vogelsberger2014introducing}.
However, at small scales \mbox{(comoving distances $\lesssim 100$ kpc)} tensions between observations and predictions from the power-law halo mass function, and power-law \shmr~relation, have had contradicting results. For example, in the Local Group, three problems were identified:
the ``missing satellites problem" \citep{ Klypin1999,Moore1999,strigari2007redefining}, the ``cusp-core" problem \citep{moore1994evidence, navarro1997universal, moore1999cold,deblok2001,kuziodenaray2008}, and ``too big to fail" \citep{boylan2011too}.
Several solutions to these challenges have been proposed, which include modifying the $\Lambda$CDM model to change the inflationary power spectrum \citep{kamionkowski2000,zentner2002} or the dark-matter model \citep{colombi1996,hu2000,kaplinghat2000,spergel2000,bode2001}, as well as including baryonic physics into $\Lambda$CDM simulations \citep{read2005,read2019,navarro1996cores,benson2002,busha2010impact,donghia2010,zolotov2012,brooks2013,dutton2016,sawala2016apostle,wetzel2016reconciling,bose2018,simpson2018,zavala2019,richings2020,samuel2020}. Through meticulous effort there has been considerable progress in addressing these small-scale problems, both observationally and theoretically, in the Local Group. Seeking data beyond our immediate neighborhood is the next step for understanding galaxy evolution, to ascertain that solutions to small-scale structure problems are not overtuned to the Milky Way.

For instance, the ``missing satellites problem", the 2000-era observation that the known number of Milky Way satellites was far lower than the number of predicted satellite galaxies---based on the abundance of predicted CDM subhalos---is framed in the context of the Milky Way. Since then, the bright satellite population of the Milky Way has been extensively studied, and is now complete and matches predictions from CDM simulations with  baryons. Similarly, Milky-Way-like galaxies have also been greatly studied, and their satellite luminosity functions---along with that of the Milky Way---are a commonly used testbed for galaxy formation in the context of $\Lambda$CDM structure formation, and of dark-matter physics \citep{willman2005new,zucker2006,belokurov2008,koposov2008luminosity,koposov2018,mcconnachie2008,martin2009,walsh2009,kennedy2014,spencer2014,drlica2015eight,kim2015,Laevens15,li2016,Torrealba16b,torrealba2019,danieli2017dragonfly,geha2017saga,Jethwa:2016gra,kim2018missing,muller2019dwarf,bennet2019m101,homma2019,nadler2019,nadler2020milky,carlsten2020a,davis2020lbt,mau2020}.
Satellite galaxy counts are used as a proxy for halo counts in studies which compare
the Local Group to predictions of $\Lambda$CDM. By using galaxy formation models that include baryons in numerical simulations \citep{zolotov2012,wetzel2016reconciling,sawala2016apostle,fattahi2018,buck2019}, or performing satellite completeness corrections \citep{tollerud2008hundreds,koposov2008luminosity,hargis2014,Jethwa:2016gra,kim2018missing,newton2018total,nadler2020milky}, these studies have shown that satellites in the Milky Way are consistent with empirical models for how galaxies inhabit halos.
However, there is worry that they may be overtuned to the Milky Way and Milky-Way-like-galaxies.
Furthermore, studies of Local Group dwarf galaxies have supported the use of a power-law stellar and halo mass function \citep{drlica2015eight,dooley2017observer,kim2018missing,nadler2020milky} 
which can be related to a galaxy's satellite luminosity function through the $M_*$-$M_{\text{halo}}$ relation. 

Observationally, new Local Volume searches for faint galaxies have increased the total numbers for a variety of environments, spanning low-mass ($\sim$ Magellanic-Cloud-mass) hosts to group environments \citep{chiboucas2013,spencer2014,carlin2016,bennet2017discovery,muller2018leo, greco2018illuminating, smercina2018,crnojevic2019,byun2020, carlsten2020a,davis2020lbt,habas2020,karachentsev2020,tanoglidis2020}. These studies, going beyond the specific Milky-Way-mass hosts explored to contextualize the Milky Way's satellites, allow us to perform more thorough tests of predictions to shed light on the physical processes that govern satellite galaxy evolution as a function of environment.
However, with most of the data coming from the Local Volume (redshift of $z \sim 0)$, the physics of any environmentally driven process affecting low-mass satellites is unlikely to be fully captured, possibly resulting in biased conclusions
\citep[see][for discussions of how typical the Local Volume is]{besla2018frequency,neuzil2020}. Therefore it is necessary to study these systems outside of the Local Volume, and push the limit to lower-mass hosts than have typically been studied before (i.e., less massive than the Milky Way), larger volumes, and higher redshifts to understand their hierarchical structure formation as a function of time.

Searching for faint satellites around more distant hosts is especially challenging, both because of their faintness and low surface brightness.
However, recent studies use statistical approaches to look for an overdensity of galaxies near a host, relative to the background, in order to measure the satellite abundance, spatial distribution, and to compare with galaxy formation models 
\citep{wang2010distribution,guo2011,Nierenberg2011,Nierenberg2012,nierenberg2013cosmic, Nierenberg2016,Sales2013,tal2014cookie, xi2018quantifying,tinker2019probing}. 
These statistical approaches reveal that we can learn about population-level properties of satellite galaxies while having only probabilistic knowledge whether a given object is a satellite of a background/foreground galaxy.

The shape and amplitude of the faint end of the satellite luminosity function---as a function of redshift and host stellar mass---can be used to understand the efficiency of star formation, while providing constraints on $\Lambda$CDM. At low redshift, for example, \citet{Sales2013} compared the satellite luminosity function of host galaxies spanning the range $7.5 \leq \log(M_*/M_{\odot}) \leq 11.0$ from the Sloan Digital Sky Survey \citep[SDSS;][]{abazajian2009seventh}, finding that the $M_*-M_{\text{halo}}$ relation and the abundance of satellites can be well constrained at the low-mass end, and is in agreement with predictions within $\Lambda$CDM.
At higher redshifts $(0.8 < z < 1.5)$, \citet{Nierenberg2016} extended the measurements of the satellite luminosity function to fainter satellites, 
from the Cosmic Assembly Near-infrared Deep Extragalactic Legacy Survey  \citep[CANDELS;][]{koekemoer2011candels} to obtain a time evolved luminosity function.
By using a Bayesian statistical method, they were able to infer parameters of the satellite luminosity function for hosts with stellar mass between $10.5 < \log(M_*/M_{\odot}) < 11.5$. 
Their findings showed that the number of satellite galaxies around hosts at high redshift and higher stellar mass, were underestimated by models that accurately predicted them for Milky Way mass hosts at low redshifts.

In this work, we use a Bayesian statistical approach, based on the \citet{Nierenberg2016} method, to measure the satellite luminosity function, as a function of redshift, for faint satellites $m_{f814W} < 25$, at redshifts up to almost half the age of the universe, $0.1 < z < 0.8$.
The host galaxy sample has a stellar mass range between the Large Magellanic Cloud (LMC) and the Milky Way,
$9.5<\log(M_*/M_{\odot})<10.5$, and comes from the Cosmic Evolution survey  \citep[COSMOS;][]{scoville2007cosmic}. Exploring this new host stellar mass regime at high redshift allows us to further study galaxy hierarchical structure formation.
We compare the shape, amplitude, and redshift evolution of the Cumulative luminosity function (CLF) to results from \citet{Sales2013} at lower redshift and \citet{Nierenberg2016} at higher redshift,
making this the first study of satellite systems of low-mass hosts across cosmic time. Additionally, we also use the characteristics of the luminosity functions to constrain the lower end slope $\beta$ of the stellar mass to halo mass relation. 

This paper is divided as follows: We present the galaxy catalog used to select satellite and host galaxies in Section \ref{sec:data}. In Section \ref{sec:host_galaxy_selection} we describe how host galaxies were selected, and their properties. Section \ref{sec:object_detection} explains how objects surrounding our selected hosts were detected, and Section \ref{sec:radial_dist} details their expected radial distribution. We carefully review the methodology of the Bayesian statistical analysis performed on the data in Section \ref{sec:modeling populations}, based on \citet{Nierenberg2016}, which allows us to detect a satellite overdensity signal above that of the background/foreground objects. Section \ref{sec:Results} presents the results of our statistical model and describes the measured cumulative satellite luminosity function. We present the satellite luminosity function and implications for $\Lambda$CDM, and the $M_*$--$M_{\text{halo}}$ models in Section \ref{sec:Discussion} along with their broader implications.  We summarize our key findings in Section \ref{sec:Summary}.

\begin{figure*}
    \includegraphics[width=1\linewidth]{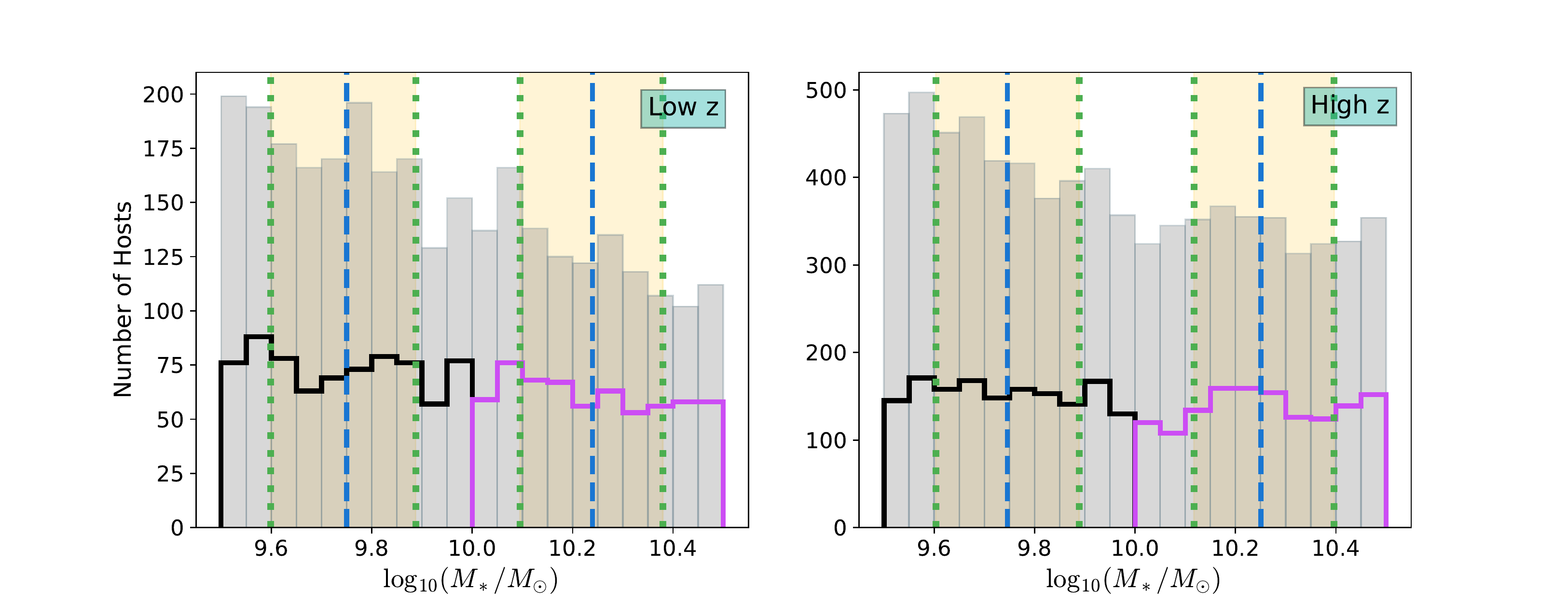}
    \caption{The stellar mass distribution of the spectroscopic low-mass hosts (black outline) and high-mass hosts (magenta outline) over-layed on the initial photometric redshift sample selected from COSMOS (gray), divided into \textit{left:} low-redshift bin $0.1 < z < 0.4$, and \textit{right:} high-redshift bin $0.4 < z < 0.8$. The yellow shaded regions represents the standard deviation from the dashed line median of each data set. The medians, in $\log\left(M_*/M_{\odot}\right)$, measured for the low-mass, low-$z$; high-mass, low-$z$; low-mass, high-$z$; and high-mass, high-$z$ sample are: 9.7, 10.2, 9.7, 10.3 respectively.}
    \label{fig:host_mass_dist}
\end{figure*}

\section{Data}
\label{sec:data}

To study the satellite luminosity function we select hosts and satellite candidates from the Cosmic Evolution Survey (COSMOS)\footnote{The COSMOS image cutouts are available at https://irsa.ipac.caltech.edu/data/COSMOS};
a 2-degree field surveyed using the Hubble Space Telescope's Advanced Camera for Surveys (ACS) Wide Field Channel  \citep[WFC;][]{koekemoer2007cosmos}. The high-resolution imaging for this survey was obtained with a single-orbit ACS I-band F814W exposures \citep{scoville2007cosmos}, with
a photometric $3 \sigma$ point source depth of 26.2 \citep{laigle2016cosmos2015}.
This wide imaged area and deep photometric data allows us to achieve a rigorous study of satellites like Fornax ($M_V \sim -13$) that are 5 magnitudes fainter than LMC luminosity hosts \citep[$M_V \sim -18$;][]{mcconnachie2012observed} out to a redshift of 0.8.

We benefit from having multi-band photometry and spectroscopy, the latter via the zCOSMOS\footnote{``zCOSMOS-bright" 20k spectroscopic redshift catalog is available at http://cesam.lam.fr/zCosmos} survey \citep{lilly2007zcosmos, lilly2009zcosmos} which contains spectra for approximately 20,000 $I$-band objects at redshifts $z \leq 1$.
However, the satellites in this study are too faint for efficient and complete spectroscopic follow-up, so we must identify them statistically.
Additionally, the estimates of stellar mass we use are based 
on a spectral energy distribution (SED) fitting technique by \citet{bolzonella2010tracking}, where the SEDs were obtained from optical to near-infrared photometry. The uncertainty in the stellar masses was   $\sigma_{\log \mathcal{M}} \simeq 0.20$ which is smaller than the width of our stellar mass bins, described in Section \ref{sec:host_galaxy_selection}; therefore, we do not include possible effects from stellar mass uncertainties in our analysis.

\section{Host Galaxy Selection}
\label{sec:host_galaxy_selection}

Our host galaxy sample was selected with the goal of studying satellite systems of low-mass hosts over time. Therefore we chose hosts with stellar masses similar to and lower than the Milky Way $(\sim 6 \times 10^{10} M_{\odot})$, but larger than the Large Magellanic Cloud (LMC;   $\sim 10^{9} M_{\odot})$. 
In order to reduce systematic uncertainty in the host stellar mass estimate and the luminosity-function measurement, as well as to more cleanly investigate the redshift evolution of the luminosity function, we only considered hosts with spectroscopic redshifts.

The upper bound on our host stellar mass complements the previous work performed by \cite{Nierenberg2012}, who used $M_* = 10^{10.5} M_{\odot}$ as their lower bound.
The lower value chosen was $M_* = 10^{9.5} M_{\odot}$ 
to ensure we could detect satellites up to 2 magnitudes fainter than the faintest hosts in our study.  

To explore the trends between satellite populations and host mass with redshift, we divided our host galaxies into two separate stellar mass bins: \mbox{$9.5<\log_{10}(M_*/M_{\odot})<10.0$} and \mbox{$10<\log_{10}(M_*/M_{\odot})<10.5$}.
These two mass-selected host sample bins 
allows us to study how the satellite properties vary with host environment, and let us more easily compare our results with simulations matching the stellar mass/luminosity function of galaxies to their halos, and other observational results.
We also divided our hosts' spectroscopic redshifts of range $0.1<z<0.8$, into two different `low' and `high' redshift bins: $0.1<z<0.4$ and $0.4<z<0.8$; allowing us to study the satellite population evolution with respect to time. 

Figure \ref{fig:host_mass_dist} shows the host stellar mass distribution of the full photometric sample and of the selected hosts with spectroscopic redshift measurements in the two different mass and redshift bins described. 
We note that the slope of the mass function is significantly different between the two samples.
The full photometric sample has nearly twice as many low-mass hosts as high-mass hosts, whereas the spectroscopic sample we use has a roughly uniform distribution in the log of the stellar mass. Therefore, we note that by only selecting hosts with spectroscopic redshifts, our sample becomes biased toward high-mass hosts, which may be due to low-mass hosts being fainter on average and thus having fewer spectroscopic redshift measurements.

We additionally point out that by only using hosts with spectroscopic redshift measurements, our host galaxy sample becomes more complete toward blue star forming galaxies---see Table 4. of \citet{bolzonella2010tracking}. We do not expect this to significantly bias our results compared to a pure mass-selected sample because in the mass range of our hosts, the majority of isolated galaxies are observed to be blue (see e.g. Figure 5 of \citet{drory2009bimodal})

A visual inspection of all the selected galaxies was performed to exclude any hosts suffering from environmental effects such as possible merging with other galaxies.
Throughout this inspection we also classified the morphology of each host as either elliptical, spiral, or irregular so that in future work a relationship between morphology and satellite number count can be studied.

\begin{figure*}
    \centering
    \includegraphics[width=1\linewidth]{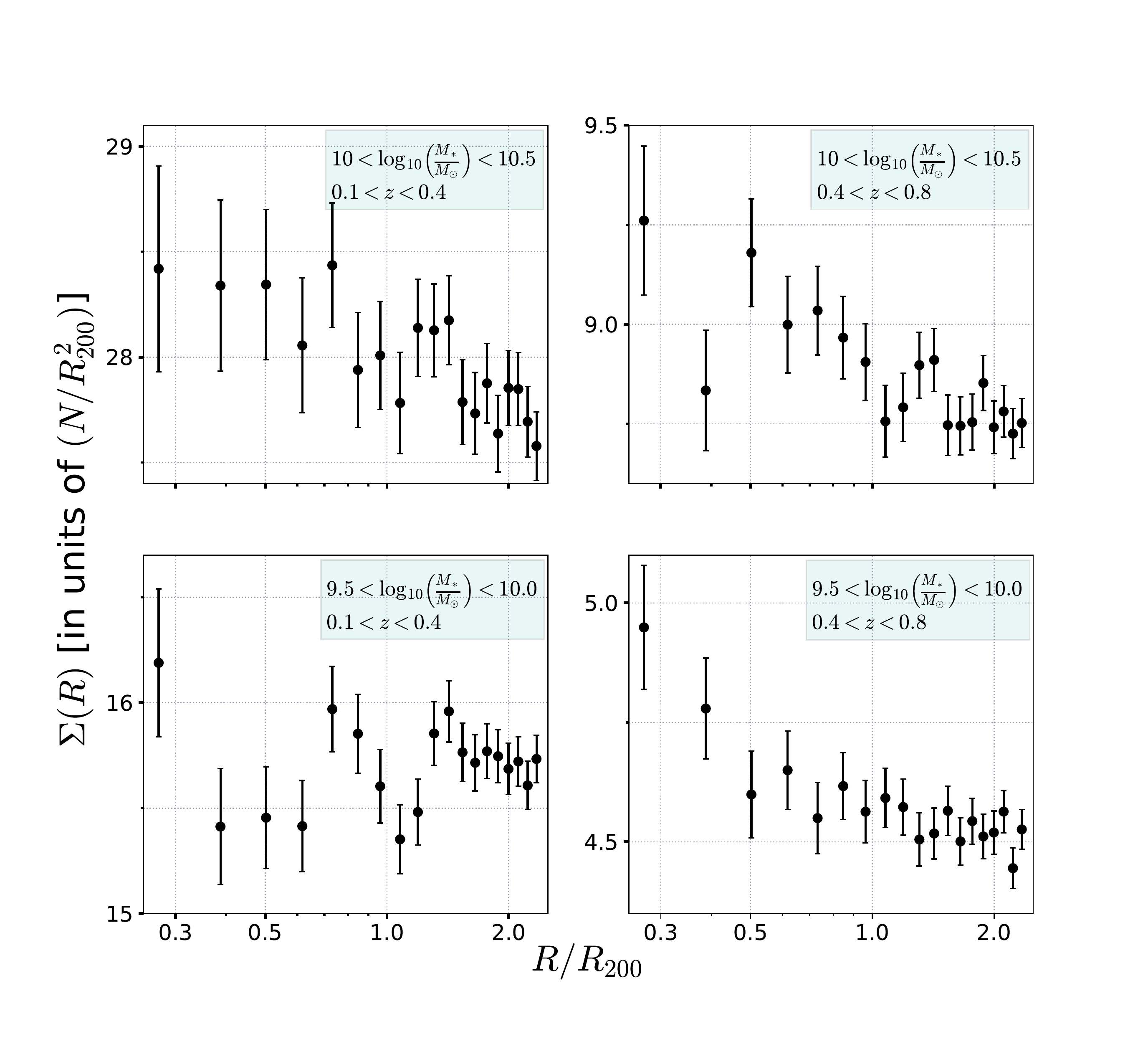}
    \caption{The average number density of objects normalized by the area of an annulus around hosts of each data set, brighter than the survey limit, as a function of radial distance in units of $R_{200}$.
    The highlighted box in each figure shows the stellar mass and redshift range of each data set. The number density of objects for the left panel is significantly higher than the right panel due to the angular size differences in a galaxy's virial radius at different redshifts. Furthermore, every point is scaled by the virial radius of each host.}
    \label{fig:radial_dist}
\end{figure*}

We required that no host galaxy be within twice the virial radius of a more massive neighboring galaxy at the same redshift 
$\left((z - z_{\text{host}})/z_{\text{host}} < 0.007\right)$.
This was done to ensure our hosts were not satellites of larger galaxies, and to not count the more massive galaxy's objects as satellites of our isolated hosts. The virial radius for each galaxy was estimated from the \shmr~relation from \cite{dutton2010kinematic}. 

Finally, we also removed any host that was less than $2.5$ times its virial radius from the edge of the COSMOS footprint. This was done in order to make sure we had radial completeness in the regions in which we would be measuring the background and satellite density of each host. 
In total, we ended up discarding a small fraction of our galaxies,
leaving us with a total of $4264$ hosts in our sample. Table \ref{tab:host_sample} shows how our total sample is divided into each of the two host stellar-mass ($9.5<\log_{10}(M_*/M_{\odot}) <10.0$ and $10<\log_{10}(M_*/M_{\odot})<10.5$) and redshift bins ($0.1<z<0.4$ and $0.4<z<0.8$), allowing us to study the satellite luminosity functions and evolution of hosts similar in size to the LMC up to the Milky Way galaxy.

\section{Object Detection}
\label{sec:object_detection}

We use the ACS I-band Photometric Catalog in order to select
objects with 'TYPE' parameter values of 1, corresponding to galaxies, down to a magnitude limit of \mbox{$m_{\text{max,survey}} = 25$}. This magnitude limit was chosen based on tests performed by \citet{Nierenberg2011} on how well satellite galaxies could be recovered as a function of their magnitude.
Because the typical photometric error for point-source objects is below 5\% \citep{laigle2016cosmos2015}, and the $3\sigma$ point-source depth is 2 magnitudes above our magnitude limit, we treat each of the satellite galaxy magnitudes as a delta function since the photometric errors are minimal for this magnitude range.

We also restrict satellite magnitudes to be \mbox{$m_{\text{min, sat}} > 18$} to reduce contamination from Milky Way stars which have a significantly increased number density relative to galaxies at magnitudes brighter than this.

To account for the difficulty of identifying faint satellites near bright hosts, we visually examined the F814W COSMOS image for each selected host galaxy, and determined a minimum radius, $R_{\text{exclude}}$, based on the simulations of \citet{Nierenberg2012} to ensure accurate photometric completeness and the exclusion of extended morphological features such as spiral arms.
A total of 212, 43, 307, and 380 galaxies were removed from our low-mass $(9.5 < \log(M_*/M_{\odot}) < 10.0)$, low-z $(0.1 < z < 0.4)$; high-mass $(10.0 < \log(M_*/M_{\odot}) < 10.5)$, low-z; low-mass, high-z $(0.4 < z < 0.8)$; and high-mass, high-z samples respectively. 
During the visual inspection we found that the apparent sizes of some of the host galaxies were unusually large relative to the virial radius, resulting in a large $R_{\text{exclude}}$, which we attribute to some type of catastrophic failure in their stellar mass measurement. Therefore, to prevent excessive exclusion of the hosts' virial radii, we removed all host galaxies with $R_{\text{exclude}} > 0.35 R_{200}$. 
Finally, we also thoroughly inspected the data to find any outlier host that could skew the inference. This meant looking for hosts with an unusual number of objects around them, or with an unusually high number of objects per angular area. 

In the following sections, we will describe the analysis performed on the host and object data we have just discussed. We will begin by looking into the radial distribution of the satellites around each host (Sec.~\ref{sec:radial_dist}), and follow with the statistical modeling we employ on our data (Sec.~\ref{sec:modeling populations}).

\section{Radial Distribution}
\label{sec:radial_dist}

One way to detect a population of satellites is by studying the mean number density of objects in concentric annuli as a function of distance from the hosts. The satellite galaxy signal should appear as a rising power law toward the host, relative to the background galaxy density beyond the virial radius.
In the case of a null satellite detection, the surface density of satellites should be roughly constant everywhere, due to the approximate isotropic and homogeneous distribution of the background/foreground galaxies.

Figure~\ref{fig:radial_dist} shows the average number density of objects per unit area, as a function of radial distance from the halo centers, with distances scaled by the virial radius of each host.

When we compare similar stellar masses in the low and high redshift bins, corresponding to the left and right columns of Figure~\ref{fig:radial_dist} respectively, we see that the number density of objects is greater for the lower redshift range. This is due to the larger angular size of the galaxy's virial radius, causing there to be a larger number of projected background/foreground objects per unit area scaled by virial radius at lower redshift. 

For all hosts with stellar masses \mbox{$10.0 < \log(M_*/M_{\odot}) < 10.5$}, as well as the high-redshift hosts with stellar mass $9.5 < \log(M_*/M_{\odot}) < 10.0$, we see a clear overdensity of objects near the center of the halo, signifying the presence of satellites.
Additionally, the signal plateaus at large radii where there are few or no satellites, as expected. 
On the other hand, for hosts with stellar masses $9.5 < \log(M_*/M_{\odot}) < 10.0$ in the low redshift bin, we do not see the previously described trend near the center of the halo.
This suggests that we do not have a large enough sample of hosts to detect a signal above the background, therefore we expect to obtain an upper limit on the satellite luminosity function for this data set.

\section{Statistical Modeling of the satellite and background galaxy populations}
\label{sec:modeling populations}

\begin{table*}
	\centering
	\caption{Parameter definitions and Priors used for the satellite and background model. (a) The Uniform prior is set between minimum and maximum values (min,max). (b) The Gaussian prior is defined by the mean and standard deviation (mean,std). For the background model, the mean and standard deviation were chosen from background/foreground density measurements. While in the satellite model, the Gaussian parameters were selected based on previous studies.}
	\label{tab:parameters}
	\begin{tabular}{l l l}
    	\hline
    	\hline
    	Parameter & Description & Prior \\
    	\hline
    	\hline
    	& Satellite Model & \\
    	\hline
    	$N_{s}$ & Total number of satellites per host between $\Delta m_{\text{ideal}}^{\text{min}} < \Delta m < \Delta m_{\text{ideal}}^{\text{max}}$ and $0.07 < R/R_{200} < 0.5$ & Uniform$(0,20)^{\text{a}}$ \\
    	$\alpha_s$ & Faint-end slope of the satellite luminosity function & Uniform$(-2.9,0)$ \\
    	$\delta_{m,\text{o}}$ & Bright-end cutoff of the satellite luminosity function & Uniform$(-8,4)$ \\
    	$\gamma_p$ & Logarithmic slope of the satellite radial distribution & Gaussian$(-1.1,0.3)^{\text{b}}$ \\
    	\hline
    	& Background Model & \\
    	\hline
    	$\Sigma_{b,\text{o}}$ & Number of all background/foreground objects per arcmin$^2$ with $I_{814} < 25$ & Gaussian$(45,0.1)$ \\
    	$\alpha_b$ & Slope of the background/foreground object luminosity function & Gaussian$(0.3,0.001)$ \\
    	\hline
    	\hline
	\end{tabular}
\end{table*}

In this section, we describe how we model the satellite and background galaxy populations taking into account the properties of each host galaxy. 
Recall that we previously showed with Figure \ref{fig:radial_dist} that the number density of satellite galaxies increases towards the host, while the background/foreground galaxies have a homogeneous and isotropic number density signal. Therefore, by inferring the properties of this combined signal, and using prior information about the background/foreground objects, the satellite number density signal can be isolated.

We start the following subsections by describing the Bayesian statistical model used to infer the parameters of the satellite and background/foreground galaxies in \S\ref{subsec: modeling sats and background galaxies}. A detailed description of each of the distributions follows, starting with the radial distribution in \S\ref{subsec: New rad dist}, the luminosity distribution in \S\ref{subsec: New lumi dist}, and the object number distribution in \S\ref{subsec: New num dist}.

\subsection{Statistical Analysis}
\label{subsec: modeling sats and background galaxies}

Three different properties were inferred using a Bayesian statistical model: 1) The probability an object is a background/foreground object or satellite, 2) The luminosity functions for satellites and background/foreground objects and 3) the radial distributions for satellites and background/foreground objects. By assuming each of these properties are separable, we find the probability distribution function (PDF) of the parameters $\bm{\theta}$ in each distribution with the model described bellow.

Our complete data sample, $\mathbf{D}$, is made up of every host and its surrounding objects, $\mathbf{D} = \{\mathbf{D}_{j=1}, \mathbf{D}_{j=2},\dots, \mathbf{D}_{j=N}\}$. Each host system, $\mathbf{D}_j$, has measurements of the host's magnitude, given by $\mathbf{h}_j$,
and the number of objects observed for that host $N_j^{\text{obs}}$, along with the positions $x_{N_j^{\text{obs}}}$ and magnitudes $m_{N_j^{\text{obs}}}$ of those objects given in $\mathbf{d}_j = \{N_j^{\text{obs}}, \{x_1, x_2, \dots, x_{N_j^{\text{obs}}} \}, \{m_1, m_2, \dots m_{N_j^{\text{obs}}} \} \}$. Therefore, each host has data $\mathbf{D}_j = \{ \mathbf{h}_j, \mathbf{d}_j \}$. To find the PDF of the model parameters $\bm{\theta}$, we use Bayes' theorem given by
\begin{align}
    \text{Pr}(\bm{\theta}|\mathbf{D}) \propto \text{Pr}(\mathbf{D}|\bm{\theta}) \text{Pr}(\bm{\theta})\ .
\end{align}
The first term in this equation, $\text{Pr}(\mathbf{D}|\bm{\theta})$
is known as the likelihood function, and the second term, $\text{Pr}(\bm{\theta})$, is the prior knowledge of the parameters.
There are two different sets of parameters inferred from the model:
the satellite, \mbox{$\bm{\theta}_s =\{N_{s}, \alpha_s, \delta_{m,\text{o}}, \gamma_p\}$}, and background/foreground model parameters, \mbox{$\bm{\theta}_b =\{\Sigma_{b,o}, \alpha_b\}$}. A description for each model parameter is shown in Table \ref{tab:parameters}, and further described in \S \ref{subsec: New rad dist}, \ref{subsec: New lumi dist}, and \ref{subsec: New num dist}.

For each parameter, we assign either a uniform of gaussian prior,  $\text{Pr}(\bm{\theta})$, based on our previous knowledge of the parameter. 
The limits and values for these distributions can be found in Table \ref{tab:parameters}.

The likelihood function can be re-written as
\begin{align}
\label{eq: likelihod each host}
    \text{Pr}(\mathbf{D}|\bm{\theta}) = \prod_{j=1}^{N_{\text{host}}} \text{Pr}(\mathbf{d}_j|\mathbf{\bm{\theta},\mathbf{h}}_j),
\end{align}
showing that it is composed of the product of each individual host galaxy's likelihood function.
For each host $j$, the likelihood is separable between the probability of measuring the total number of objects around a host
$\text{Pr}(N_j^{\text{tot}}|\bm{\theta})$, with magnitudes between \mbox{$m_{\text{sat}}^{\text{min}} < m < m_{\text{survey}}$}, and the position of each object with a given luminosity $\text{Pr}({R_i}, \Delta m_i|\bm{\theta}, \mathbf{h}_j)$, 
given the model parameters $\bm{\theta}$. This is due to the total number of objects being independent of their distribution around the host. The full expression of the likelihood for the objects of a host can be written as
\begin{align}
    \text{Pr}(\mathbf{d}_j|\bm{\theta},\mathbf{h}_j) = \text{Pr}(N_j^{\text{tot}}|\bm{\theta})  \prod_{i=1}^{N_j^{\text{obs}}} \text{Pr}({R}_i, \Delta m_i|\bm{\theta}, \mathbf{h}_j) ,
    \label{eq: host individual likelihood}
\end{align}
where $i$ represents the object around the host.
The term $\text{Pr}({R}_i, \Delta m_i|\bm{\theta}, \mathbf{h}_j)$ depends on the sum of a satellite's probability of existing at a location $R_i$ with a difference in magnitude from the host $\Delta m_i$, and the probability a background/foreground object exists at that same location with the same magnitude difference. Because an object's position is independent of its magnitude, we further separate these two probabilities and express the sum as 
\begin{align}
\begin{split}
    \text{Pr}({R}_i,\Delta m_i|\bm{\theta}, \mathbf{h}_j) =  \text{Pr}({R}_i|\bm{\theta}, \mathbf{h}_j, S) 
    \text{Pr}(\Delta m_i|\bm{\theta}, \mathbf{h}_j, S) \text{Pr}(S|\bm{\theta},\mathbf{h}_j) \\
    + \text{Pr}({R}_i|\bm{\theta}, \mathbf{h}_j, B) 
    \text{Pr}(m_{\text{back}}|\bm{\theta}, B) \text{Pr}(B|\bm{\theta},\mathbf{h}_j) \ ,
    \label{eq:location and lumi likelihood}
\end{split}{}
\end{align}
where the terms $\text{Pr}(S|\bm{\theta},\mathbf{h}_j)$ and $\text{Pr}(B|\bm{\theta},\mathbf{h}_j)$ refer to the relative probability an object is a satellite or a background/foreground galaxy respectively, and $m_{\text{back}}$ is the magnitude of the background/foreground object.
The PDF of the parameters $\bm{\theta}$ in each distribution
was computed separately for the four data sets using a Markov Chain Monte Carlo (MCMC) method. To guarantee convergence, at least $10^4$ iterations per chain were performed based on the results from \cite{Nierenberg2012}. 
We also verified the Bayesian statistical code, used to infer the model parameters, by running the COSMOS photometric catalog used in \citet{Nierenberg2012}, and reproducing their results for the stellar mass range \mbox{$10.5 < \log( M_{*,\text{host}}/M_{\odot}) < 11.5$} at redshifts $0.1 < z < 0.8$.

The following subsections describe the probability functions in detail.

%
%

\subsection{Radial Probability Distribution}
\label{subsec: New rad dist}
Observationally and in simulations, satellites with stellar masses above Fornax-like objects (\mbox{$\sim 10^8 M_{\odot}$)} of low-mass galaxies, have radial distributions which are independent of both satellite color and luminosity \citep{guo2013spatial,wang2014satellite, sales2015colours, prescott2011galaxy}. Thus in our analysis, we assume that the luminosity function is separable from the radial distribution.
We construct a model to find the radial probability density $\text{Pr}({R}_i|\bm{\theta}, \mathbf{h}_j)$ for satellites and background/foreground objects located at some position $R_{i}$, within an annulus of inner and outer radii $R_{\text{exclude}}$ and \mbox{$R_{\text{max,ideal}} = 0.5 R_{200}$} respectively. The lower limit represents the minimum radius of each host where we can accurately detect satellites, and the upper limit corresponds to where the satellite signal becomes significantly noticeable above the background, based on Figure \ref{fig:radial_dist}. In the following subsections we describe the details of the model for satellites and background/foreground objects.

\subsubsection{Satellites}
\label{subsubsec: radial dist satellites}

The satellite population is characterized by the radial distribution trend seen in Figure~\ref{fig:radial_dist}. The 
increase in the surface density of objects toward the host galaxies centers, observed in three of our four host samples, is a clear sign of satellite galaxy detections. We model the overdensity signal as a power-law 
\begin{align}
   \Sigma_s(R) \propto R^{\gamma_p},
   \label{eq:radial powerlaw}
\end{align}
and used a well-motivated Gaussian prior for $\gamma_p$, with mean $-1.1$ and standard deviation $0.3$ (see Table \ref{tab:parameters}). These values were chosen from several slope measurements \citep{Nierenberg2011,tollerud2011small, watson2012extreme} of satellites with different morphologies, magnitudes, and host masses.

To calculate the probability density of a satellite galaxy being located at some position $R_i$, shown in the first term of Equation~\ref{eq:location and lumi likelihood}, we normalize the satellite number density to obtain:
\begin{align}
    \text{Pr}({R_i}|\bm{\theta},\mathbf{h}_j, S) = \left( \frac{\bm{\gamma}_p + 2}{R_{\text{max,ideal}}^{\bm{\gamma}_p +2} - R_{\text{exclude},j}^{\bm{\gamma}_p +2}} \right) \frac{R_{i}^{\bm{\gamma}_p +1}}{2 \pi},
\end{align}
where the radial limits are also \mbox{$R_{\text{max,ideal}}$} and \mbox{$R_{\text{exclude}}$}.

\subsubsection{Background/Foreground objects}
\label{subsubsec: radial dist background}

The spatial distribution of background/foreground objects is modeled to be homogeneous and isotropic, with a number density $\Sigma_{b,\text{o}}$ for all objects with magnitudes brighter than the survey magnitude limit around each host $j$. To create a prior for $\Sigma_{b,\text{o}}$, we measured the galaxy number counts between $R/R_{200} = 1.5 - 2.0$ based on our observations from Figure \ref{fig:radial_dist}, where the background signal becomes roughly constant. These priors were measured separately for each host and stellar mass bin to account for any differences in line of sight structures. However we found that the background densities were consistent across all samples.

The probability density of finding a background/foreground object at some position $R_{i}$ by
\begin{align}
   \text{Pr}({R}_i|\bm{\theta}, \mathbf{h}_j, B) =
    \frac{R_{i}}{A_j},
\end{align}
where the area $A_j$ is the annulus between $R_{\text{exclude}}$ and \mbox{$R_{\text{max,ideal}} = 0.5 R_{200}$}.

%
%

\subsection{Luminosity Function}
\label{subsec: New lumi dist}

The probability of observing a satellite with a given \mbox{$\Delta m = m_{\text{sat}} - m_{\text{host}}$}, or background/foreground object with a magnitude $m_{\text{back}}$, partly depends on the range of observable $\Delta m$ and background magnitude values for satellites and background/foreground objects respectively. 
We measure satellite luminosities relative to their host magnitudes, therefore the observable satellite region, $\Delta m_{\text{obs}}$, is defined to be between \mbox{$\Delta m_{\text{obs}}^{\text{min}} = m_{\text{sat}}^{\text{min}} - m_{\text{host}}$} and \mbox{$\Delta m_{\text{obs}}^{\text{max}} = m_{\text{sat}}^{\text{max}} - m_{\text{host}}$}, where $m_{\text{sat}}^{\text{max}} = 25$ corresponds to the survey magnitude limit and $m_{\text{sat}}^{\text{min}} = 18$ is the brightest satellite magnitude in our sample. 

Below we describe the probability density for the satellite and background/foreground luminosity functions.

\subsubsection{Satellites}
\label{subsubsec: lumi function satellites}

We model the satellite luminosity function as a Schechter function given by:
\begin{align}
\begin{split}
  \Phi(\Delta m_i)
  \propto \, 10^{0.4(\alpha_s + 1)(\delta_{m,o} - \Delta m_i)}
    &\exp{\left[-10^{0.4(\delta_{m,o} - \Delta m_i)} \right]},
\end{split}
\label{eq:schechter lf}
\end{align}
where $\Phi(\Delta m_i)$ is the number density of the satellite galaxies, as a function of $\Delta m_i$, and the slope $\alpha_s$ and turnover $\delta_{m,o}$, are left as free parameters in the fit.

The probability density for the satellite luminosity function 
is defined by
\begin{align}
\begin{split}
  \text{Pr}(\Delta m_i|\bm{\theta},\mathbf{h}_j, S) 
  = \frac{\Phi_j(\Delta m_i)}{\int_{\Delta m_{\text{obs}}^{\text{min}}}^{\Delta m_{\text{obs}}^{\text{max}}} \Phi_j(\Delta m_i) \, d(\Delta m_i)},
\end{split}
\label{eq:lumi prob dist}
\end{align}
where the Schechter function has been normalized between the observable $\Delta m_{\text{obs}}$ limits.

\subsubsection{Background/Foreground objects}

The luminosity function of background/foreground objects is represented by a power-law \citep{benitez2004faint} given by:
\begin{align}
    \Phi_b(m_{\text{back}}) \propto 10^{\alpha_b m_{\text{back}}} 
\end{align}
where the model parameter $\alpha_b$ represents the slope of the background/foreground luminosity distribution.
The luminosity model prediction for the background objects
$\text{Pr}(m_{\text{back}}|\bm{\theta}, B)$ is then given by normalizing the power-law luminosity function between $m_{\text{back}}^{\text{min}} = 18$ and $m_{\text{survey}} = 25$ resulting in:
\begin{align}
    \text{Pr}({m_{\text{back}}}|\bm{\theta}, B) = \frac{ \alpha_b \log(10) \, 10^{\alpha_b (m_{\text{back}} - m_{\text{back}}^{\text{min}})}}{10^{\alpha_b (m_{\text{survey}}-m_{\text{back}}^{\text{min}})} -1}. 
\end{align}

%
%

\subsection{Satellite and Background/Foreground Galaxy Number Distribution}
\label{subsec: New num dist}

The probability of measuring the total number of objects $N_j^{\text{tot}}$ around a host $j$ is given by the term
$\text{Pr}(N_j^{\text{tot}}|\bm{\theta)}$ 
in Equation~\ref{eq: host individual likelihood}, 
which we define by a Poisson probability.
To calculate the model predicted number of objects, we sum the model prediction for the observable number of satellites $N_{s,j}^\text{obs}$ and background/foreground galaxies $N_{b,j}^{\text{obs}}$ for each host $j$. Bellow we give the specific model for each type of galaxy population.

\begin{table*}
\centering
\caption{Satellite model parameters' median and one sigma confidence interval in the \emph{ideal} radial range $0.07 < R/R_{200} < 0.5$, also defined between the \emph{ideal} $\Delta m$ range. 
The hosts with stellar mass $9.5 < \log (M_*/M_{\odot}) < 10.0$ and redshift $0.1 < z < 0.4$ do not appear on this table due to only having upper limits. Note that these parameters are also measured outside of these \emph{ideal} ranges in order to construct the cummulative luminosity function.}
\label{tab:dms measured}
\begin{tabular}{cccccccc}
\hline
\hline
$\log \left( M_{*, \text{host}}/M_{\odot}\right)$ & $z_{\text{host}}$ & $\Delta m_{\text{ideal}}^{\text{min}}$ & $\Delta m_{\text{ideal}}^{\text{max}}$ & $N_{s}$ & $\alpha_s$ & $\delta_{m,\text{o}}$ & $\gamma_p$ \\ \hline \hline
$9.5-10.0$                                        & $0.4-0.8$                           & $2$                       & $3.5$                      & $0.15 \pm 0.04$           & $-1.2 \pm 0.5$                & $-0.1 \pm 1.8$                 & $-1.2 \pm 0.3$                     \\ 
$10.0-10.5$                                       & $0.1-0.4$                           & $2$                       & $5.5$               & $0.3 \pm 0.2$                 & $-1.2 \pm 0.2$              & $-4.1 \pm 2.3$                & $-1.1 \pm 0.3$                     \\ 
$10.0-10.5$                                       & $0.4-0.8$                           & $2$                       & $4.0$                & $0.2 \pm 0.08$                 & $-1.2 \pm 0.5$              & $-0.3 \pm 1.3$                & $-1.1 \pm 0.3$                     \\ \hline \hline
\end{tabular}
\end{table*}

\subsubsection{Satellites}
\label{subsubsec: new satellite detection}

The model prediction for the number of satellites per host parameter, $N_{s}$, is measured between \mbox{$R_{\text{min,ideal}} = 0.07 R_{200}$} and \mbox{$R_{\text{max,ideal}} = 0.5 R_{200}$}, defined within an \emph{ideal} $\Delta m$ range. The lower limit of the radial range was determined based on the previous observational analysis by \cite{Nierenberg2016}, who showed that this inner radius ensures accurate photometry of objects near the host.

The \emph{ideal} $\Delta m$ range was chosen to be where the satellite luminosity function could be measured for at least $\sim 30 \%$ of all hosts.
We set this lower limit to $\Delta m_{\text{min,ideal}} = 2$, which allowed us to observe satellites at least two magnitudes fainter than the survey magnitude limit (see Section \ref{sec:host_galaxy_selection}). The maximum \emph{ideal} limit $\Delta m_{\text{ideal}}^{\text{max}}$ varied with each data set, depending on the number of hosts that exist in the different $\Delta m$ bins above the survey limit. 
The selected \emph{ideal} $\Delta m$ values are listed in Table~\ref{tab:dms measured} for each data set. In Section~\ref{sec:Results} we will use the model predicted $N_{s}$ as a normalization factor to extend our satellite search past the \emph{ideal} radial and $\Delta m$ range, to find the CLF in the full virial radius of each host.

We infer the parameter $N_s$ based on the
ratio of satellites we can directly observe per host $N_{s,j}^{\text{obs}}$, defined between an observable radial and $\Delta m$ range, to the predicted number of satellites per host $N_s$. Defining the general radial and $\Delta m$ fractions:
\begin{align}
\begin{split}
    &f_{R}(R_{\text{min}},R_{\text{max}}) = \frac{\int_{R_{\text{min}}}^{R_{\text{max}}} \Sigma(R) R \ dR}{\int_{R_{\text{min,ideal}}}^{R_{\text{max,ideal}}} \Sigma(R) R \ dR} \\ &f_{\Delta m,j}(\Delta m_{\text{min}},\Delta m_{\text{max}}) = \frac{\int_{\Delta m_{\text{min}}}^{\Delta m_{\text{max}}} \Phi_j(\Delta m) \, d(\Delta m)}{\int_{\Delta m_{\text{ideal}}^{\text{min}}}^{\Delta m_{\text{ideal}}^{\text{max}}} \Phi_j(\Delta m) \, d(\Delta m)},
    \label{eq:general fracs}
\end{split}{}
\end{align}
where $\Sigma(R)$ is the radial satellite number density from equation \ref{eq:radial powerlaw}, and $\Phi(\Delta m)$ is the Schechter Luminosity function given by Equation~\ref{eq:schechter lf}, we can express the model prediction for the observed number of satellites per host as:
\begin{align}
\begin{split}
        N_{s,j}^{\text{obs}} = 
        N_{s} &\times  f_{R}(R_{\text{exclude}},R_{\text{max,ideal}}) \\
        &\times f_{\Delta m,j}(\Delta m_\text{obs}^{\text{min}},\Delta m_\text{obs}^{\text{max}}) \ .
    \label{eq:num sats}
\end{split}{}
\end{align}
The observable radial limits range between the host's exclusion radius, $R_{\text{exclude}}$, and the ideal outer radius of $R_{\text{max,ideal}} = 0.5 R_{200}$ which ensures a detectable satellite signal.

\subsubsection{Background/Foreground Objects}
\label{subsubsec: new back/foreground obj detection}

The model predicted number of background/foreground objects for a given host $j$, is given by
\begin{align}
N_{b,j}^{\text{obs}} = \Sigma_{b,\text{o}} A_j \ .
\label{eq:number background}
\end{align}
where $A_j$ is the area between $R_{\text{exclude}}$ and $R_{\text{max, ideal}}$.

 \subsubsection{Relative probability of being a satellite or background/foreground object}

Given the model prediction for the observed number of satellites $N_{s,j}^{\text{obs}}$ and background objects $N_{b,j}^{\text{obs}}$ from Equation \ref{eq:number background} and Equation~\ref{eq:num sats} respectively, we can calculate the relative probability an object is either a satellite, $\text{Pr}(S|\bm{\theta},\mathbf{h}_j)$, or a background/foreground object, $\text{Pr}(B|\bm{\theta},\mathbf{h}_j)$, seen in Equation~\ref{eq:location and lumi likelihood} by
\begin{align}
\begin{split}
    \text{Pr}(S|\bm{\theta},\mathbf{h}_j) = \frac{N_{s,j}^{\text{obs}}}{N_{s,j}^{\text{obs}}  + N_{b,j}^{\text{obs}}} \\ \text{Pr}(B|\bm{\theta},\mathbf{h}_j) = \frac{N_{b,j}^{\text{obs}}}{N_{s,j}^{\text{obs}} + N_{b,j}^{\text{obs}}}.
\end{split}{}
\end{align}
Using the radial, luminosity, and number distributions described in the previous sections, we were able to calculate the likelihood of observing satellites or background/foreground objects with their respective luminosity and position (Equation \ref{eq:location and lumi likelihood}), given their model parameters $\bm{\theta}$.

\begin{figure*}
    \includegraphics[width=1\linewidth]{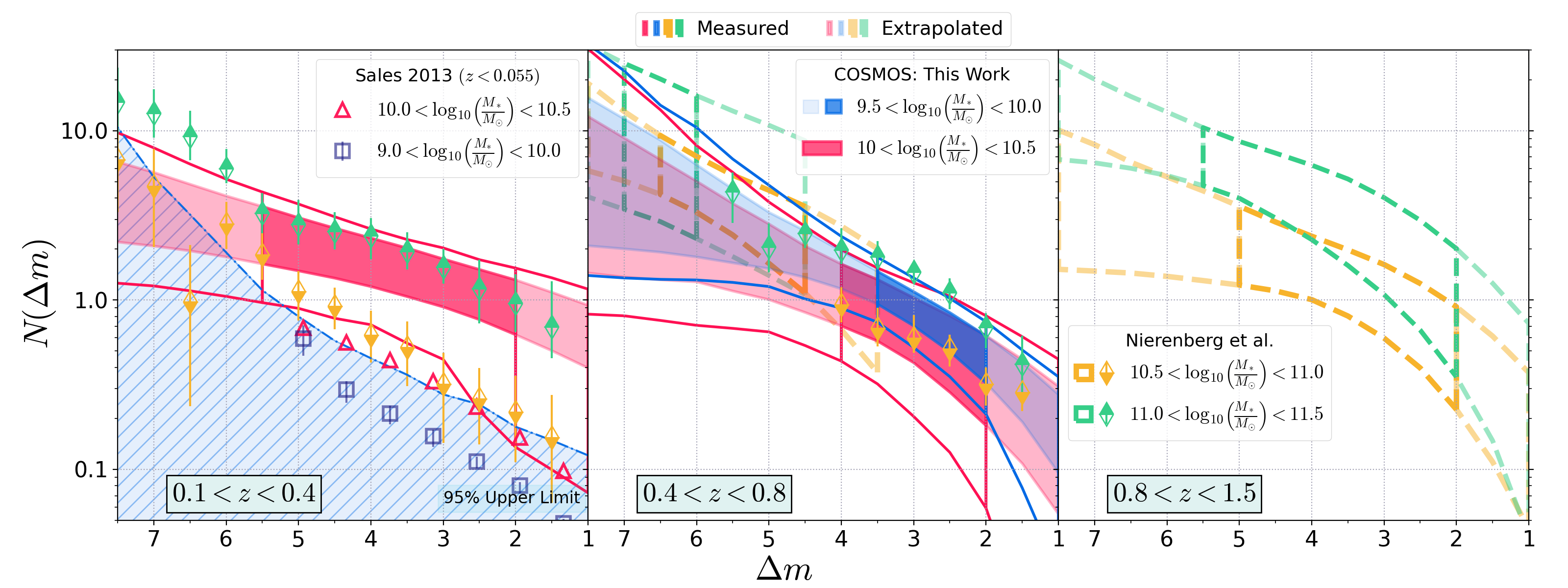}
    \caption{The cumulative satellite luminosity function (CLF) per host between $r/R_{200} = 0.07 - 1.0$ as a function of $\Delta m$, in increasing redshift bins from left to right. The dark shaded regions represent the range where we best constrained the CLF, and the lighter shaded regions show where there was less data to accurately infer the number of satellites; called the \emph{ideal} and extrapolated $\Delta m$ regions respectively. Squares, and triangles represent results from \citet{Sales2013}. The green and orange half filled diamonds and bands show the results from \citet{Nierenberg2012} and \citet{Nierenberg2016} respectively. The $1 \sigma$ confidence bands for this work are shown in pink and blue, along with the hatched blue region which represents the 95\% upper limit. We also show our $2\sigma$ confidence bands, outlined in the same pink and blue color.}
    \label{fig:luminosity_func}
\end{figure*}

\section{Results}
\label{sec:Results}

In this section, we present the CLF of the satellite population for each separate host mass and redshift bin as a function of difference in magnitude $\Delta m$.
Table \ref{tab:dms measured} lists the median and $1 \sigma$ uncertainties of the satellite model parameters for three of the data sets in the chosen \emph{ideal} $\Delta m$ range.

In Sec.~\ref{sec:modeling populations}, we set an ideal radial range for satellite searches to be between $0.07 < r/R_{200} < 0.5$ (\S \ref{subsubsec: new satellite detection}).
However we would like to extend the upper limit out to the virial radius in order to infer the total number of satellites $N(\Delta m)$ for any given $\Delta m$ bin of the CLF within the virial radius. Therefore we rescaled the model predicted number of satellites $N_{s}$ into the total number of satellites $N(\Delta m)$ using the inferred parameters and Equation~\ref{eq:num sats}, where now $R_{\text{max}} = R_{\text{virial}}$, the host's complete virial radius, and the $\Delta m$ values are chosen to encompass the entire luminosity range in intervals of $0.5$.

In order to calculate the CLF, we first drew random values from the posterior probability distribution functions and generated many luminosity functions. At a given $\Delta m$, the value of the CLF is given by the median of these functions, and the uncertainty was determined by their $1 \sigma$ and $2 \sigma$ deviation.
For the low-mass, low-z data set the uncertainty result for each parameter was large. We relate this to the fact that we have a null detection of satellites, suggesting that a larger sample of hosts is required in order to measure the satellite luminosity function for low-mass hosts at low redshift.
Because we have the best constraints in the \emph{ideal} $\Delta m$ range, the statistical uncertainties tend to be smaller there than in the extended regions at high and low $\Delta m$ values, where there exists low number of host galaxies. We call these regions of higher uncertainty the \emph{extrapolated} $N(\Delta m)$ regions.

Figure \ref{fig:luminosity_func} shows the total number of satellites per host, $N(\Delta m)$ between $0.07 < r/R_{200} < 1$, in increasing $\Delta m$ bins, which is the difference between host and satellite magnitude. Our sample with host stellar masses $9.5 < \log(M_*/M_{\odot}) < 10.0$ and $10.0 < \log(M_*/M_{\odot}) < 10.5$, blue and pink regions respectively, are shown in the two redshift bins of this work: $0.1 < z < 0.4$ and $0.4 < z < 0.8$. The darker shade of the CLF represents the \emph{ideal} $\Delta m$ range and the lighter shade of color shows the \emph{extrapolated} region.
To place our results in the context of other work, we show the SDSS/DR7 results of \cite{Sales2013}, with host stellar mass range $9.0 < \log(M_*/M_{\odot}) < 11.0 $ in the lower redshift bin.
Additionally, we include results from \cite{Nierenberg2012} and \citet{Nierenberg2016} CANDELS and COSMOS sample: $10.5 < \log(M_*/M_{\odot}) < 11.0$ and $11.0 < \log(M_*/M_{\odot}) < 11.5$ in the redshift bins of this work, along with the higher redshift bin: $0.8 < z < 1.5$ of Nierenberg et al., which also have \emph{ideal} and \emph{extrapolated} ranges.
These three increasing redshift bins, along with data at $z<0.055$, allows us to study the evolution in satellite populations and test the CDM prediction of hierarchy of structure. 
The CLF is presented out to high and low $\Delta m$ values in order to show the overall shape of the luminosity function in all the redshift bins.

\section{Discussion}
\label{sec:Discussion}

Using the CLF in Figure~\ref{fig:luminosity_func} we discuss three important features: the amplitude in section \ref{subsec: CLF}, redshift evolution in section \ref{subsec: redhift evolution}, and in section \ref{subsec: slope} the slope of the satellite luminosity functions.
We also describe a future directions for satellite galaxy studies guided by this work.

\subsection{Cumulative Luminosity Function Amplitude}
\label{subsec: CLF}

The first feature we explore is the relative amplitude of the CLF for the low- and high-mass hosts of our COSMOS data, and compare it to the amplitude results of \citet{Sales2013} and \citet{Nierenberg2016}.
We expect to see, in each redshift panel, the amplitude of the lower-mass CLF curves group together, indicating that the number of satellites for these hosts is independent of stellar mass. 
This is due to the self-similarity of the subhalo mass function \mbox{$N \propto \left(M_{\text{subhalo}}/M_{\text{halo}}\right)^{\alpha + 1}$}  \citep{giocoli2008population}; and the power-law nature of the \shmr~relation \mbox{$M_* \propto M_{\text{halo}}^{\beta}$} in the region below a characteristic halo mass peak \mbox{$M_{\text{halo}} \sim 10^{12} M_{\odot}$} \citep[e.g.,][]{leauthaud2012,behroozi2019,girelli2020stellar}. 
Above this peak, we would expect to see satellite abundances become dependent on host stellar mass \citep{nickerson2013luminosity}, and rise due to a shift in the slope of the \shmr~relation.
But for low-mass hosts, below the pivot in the \shmr~relation, i.e., the break in the power law, the number of satellites should approximately scale to the satellite and host stellar mass as $N \propto \left(M_{*,\text{sat}}/M_{*,\text{host}}\right)^{(\alpha + 1)/\beta}$, showing that for a specific difference between the host and satellite stellar mass, the number of satellites should remain constant.

We observed this trend in the center panel \mbox{$(0.4<z<0.8)$} of Figure \ref{fig:luminosity_func}, between our COSMOS hosts and those from \citet{Nierenberg2016}. The comparison shows the number of satellites increasing with increasing stellar mass, within a given $\Delta m$ bin.
The CLF bands of Nierenberg et al. show a rise in the number of satellites per host with increasing stellar mass at a $1 \sigma$ interval, due to their stellar masses being above the pivot in the \shmr~relation. This trend is also observed in the right panel of Figure \ref{fig:luminosity_func}, where we also show their \mbox{$0.8<z<1.5$} redshift bin. This increase in amplitude becomes more noticeable at higher $\Delta m$ values, where there exists an increase in the average host stellar masses. On the other hand, for smaller $\Delta m$ values, satellites and hosts are above the pivot in the \shmr~relation.  Because the \shmr~relation is well-described as a power law above the pivot, we expect the CLFs for the two host mass bins to overlap. Indeed, this is what we observe. 

Returning to the middle panel, when we reach host stellar masses from our COSMOS sample, $10.0 < \log(M_*/M_{\odot}) < 10.5$ and $9.5 < \log(M_*/M_{\odot}) < 10.0$, we see a weak dependence between stellar mass and number of satellites in those same $\Delta m$ bins.  This means that these stellar masses are in the power-law region of the stellar mass-halo mass  $(M_*$--$M_{\text{halo}})$ relation below the pivot. This trend is particularly prominent between the measured $\Delta m$ region which is shown in a darker shade.  This is consistent with \citet{leauthaud2012}, who finds that the pivot in the \shmr~relation lies at a stellar mass of $\log(M_*/M_{\odot}) = 10.8$,

The left panel of Figure \ref{fig:luminosity_func}, which corresponds to the low redshift bin $0.1<z<0.4$, shows the 95\% upper confidence limit of our low-mass $9.0 < \log(M_*/M_{\odot}) < 10.0$ hosts and the CLF of the high-mass $10.0 < \log(M_*/M_{\odot}) < 10.5$ hosts (68\% and 95\% confidence levels shown). At the $1 \sigma$ confidence interval we do not see the predicted pile-up of low-mass CLFs, especially at small $\Delta m$ values. There is, however, a small overlap at higher $\Delta m$ values which is due to the slopes of the luminosity functions being different.

In this low redshift bin, we compare the CLF amplitudes of our sample to those hosts of similar low mass, at even lower redshift $z<0.055$, from \citet{Sales2013}.
The $1\sigma$ confidence interval for all of the $\Delta m$ bins of our COSMOS hosts with stellar mass $10.0 < \log(M_*/M_{\odot}) < 10.5$ shows a slightly greater measured number of satellites than for the same stellar mass hosts from \citet{Sales2013} (pink triangles). However, our COSMOS data do overlap in the $2\sigma$ confidence interval for low values of $\Delta m$. 
For hosts with stellar mass $9.5 < \log(M_*/M_{\odot}) < 10.0$, the number of satellites in the 95\% upper limit agrees for all the data points from \citet{Sales2013}. We also see that the CLF amplitude for the higher mass range $10.5 < \log(M_*/M_{\odot}) < 11.0$ (orange half filled diamonds) of \citet{Nierenberg2016} measured fewer satellites than our lower $10.0 < \log(M_*/M_{\odot}) < 10.5$ hosts, due to the different inferred slopes. However, the two luminosity functions are consistent with each other at the 2$\sigma$ level.

Our CLFs have shown that there exists a weak dependence between host stellar mass and the number of satellites for the low-mass hosts in the COSMOS sample, especially seen in the middle panel of Figure~\ref{fig:luminosity_func}.
Although the results at low redshifts are statistically consistent with the weak trend of satellite number with host stellar mass at fixed $\Delta m$, we note that several factors could drive differences between the measured values observed in the left-hand panel of Figure~\ref{fig:luminosity_func}. 
The first is that the distribution of host stellar masses from our sample may not be the same as the other data sets we are comparing with. In Figure~\ref{fig:host_mass_dist}, we showed that our spectroscopic host sample is biased toward high stellar masses.  
This will tend to increase the number of satellites per host, relative to those of hosts from a cosmological distribution, if the host mass is near the pivot in the \shmr~relation.
We can roughly estimate how much our satellite population may be biased by assuming a similar behavior to that of figure 2. of Sales et al.. By looking at their $\log(M_*/M_\odot) = 10.25$ and $\log(M_*/M_\odot) = 10.75$ stellar mass hosts, which are of similar mass range as ours, and assuming no redshift evolution we might expect our satellite population to be biased by a factor of two. This difference would explain why our CLFs appear to be marginally higher than expected, yet still consistent with results from Sales et al., and Nierenberg et al., given our uncertainties. 

The second hypothesis for the CLF discrepancies centers on the issue of background/foreground contamination due to the presence of interlopers, which are background/foreground objects that falsely correspond with the host. 
In contrast to Sales et al., we measured interlopers as part of our model. Sales et al. used distance and velocity cuts as a way to remove interlopers, a method based on \citet{van2004probing}. Both these methods are able to produce satellite samples with low number of interlopers. However, if we only use the satellites' spectroscopic redshifts to estimate their association with hosts, we would expect to see up to $\sim 40 \%$ interloper contamination in a satellite sample \citep{besla2018frequency}. With this in mind, if our satellite sample were to have a large fraction of interlopers, then the measured number of satellites per host would decrease. However, our results, along with those of Nierenberg et al. and Sales et al., do not seem to have a significant enough interloper contamination. Therefore, the CLF discrepancy would not be affected.  

The final hypothesis to explain the disparity of the CLFs in the low $z$ bin of Figure~\ref{fig:luminosity_func}, is using the redshift dependence of the \shmr~relation. According to figure 5. of \citet{moster2013galactic}, at redshifts near $z \sim 0$ the stellar mass pivot point of the \shmr~relation is significantly lower than at redshifts $z > 0.5$. Thus, we should not expect our high stellar mass hosts ($10.0 < \log(M_*/M_{\odot}) < 10.5$) to completely overlap the CLF's of the low-mass hosts in the low-z bin. This is due to our high-mass hosts not entirely residing in the power law region, and instead being located much closer to the pivot in the \shmr~relation. 
Sales et al. also see a lack of dependence between stellar mass and satellite abundance below $\log(M_*/M_{\odot}) = 10.25$, but they do for larger stellar masses, further indicating that we should not expect a clean overlap between our low-and-high mass bins. On the other hand, because the pivot in the \shmr~relation is higher for redshifts $z > 0.5$, the overlap we see in the middle panel of Figure~\ref{fig:luminosity_func} for our low-and-high mass bins is expected.

In summary, our measurements of the CLF are consistent at the 2$\sigma$ level with other published data and with the hypothesis that CLFs should become independent of host stellar mass for low-mass hosts if expressed in terms of $\Delta m$.

\begin{figure*}
    \centering
    \includegraphics[width=1\linewidth]{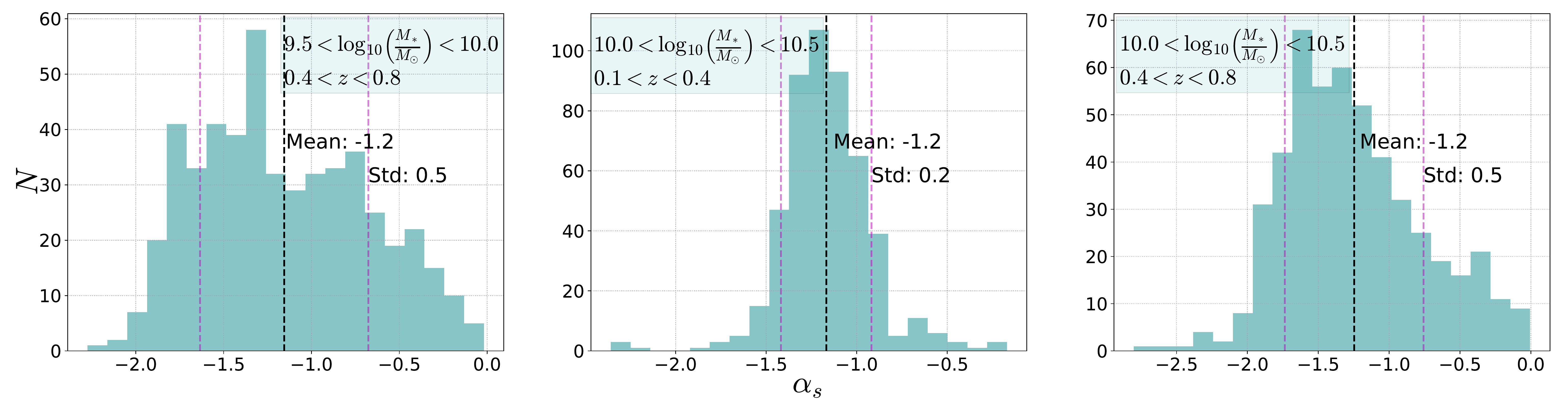}
    \caption{The posterior PDF of the faint-end slope of the satellite luminosity function, $\alpha_s$, for satellites (from left to right) of low-mass, high-$z$; high-mass, low-$z$; and high-mass, high-$z$. The mean and standard deviation are shown in each panel in black and pink, respectively.}
    \label{fig:alpha_s}
\end{figure*}

\subsection{Redshift Evolution}
\label{subsec: redhift evolution}

The second feature we describe in Figure~\ref{fig:luminosity_func} is the redshift evolution of the satellite luminosity function. 
In general we see that the satellite abundance per host for the $10.0 < \log(M_*/M_{\odot}) < 10.5$ and the $9.5 < \log(M_*/M_{\odot}) < 10.0$ COSMOS host samples remains constant as a function of redshift given the measured uncertainties.
This result is consistent with other low-redshift studies  \citep[$z<0.1$;][]{danieli2017dragonfly, nadler2019modeling} and with previous results, although for a higher stellar mass host sample.
Our results are also in agreement with the prediction from \citet{marmol-queralto2012}, \citet{tal2014cookie}, and \citet{rafieferantsoa2018}, who showed that, in the context of $\Lambda$CDM, the satellite luminosity function  should remain constant out to $z \sim 1$ for fixed host stellar mass.

\subsection{Slope}
\label{subsec: slope}

Lastly, the faint-end slope of the satellite luminosity function $\alpha_s$ can be used to infer the low-end slope of the \mbox{$M_* - M_{\text{halo}}$} relation $\beta$, placing our results in context of $\Lambda$CDM.
We perform an order-of-magnitude calculation that will help constrain this slope between \mbox{$9.5 < \log(M_*/M_{\odot}) < 10.5$} to further understand the power-law nature of the \shmr~relation.

The differential number of dark matter subhalos $dN_{\text{sub}}$ hosted by a halo of mass $M_{\text{host}}$ is given by the subhalo mass function $dN_{\text{sub}}/dM_{\text{sub}} \propto M_{\text{sub}}^{\alpha}$, with $\alpha \sim -1.9$ \citep{Springel_2008,Dooley2014, zavala2019dark}. The slope is nearly independent of whether $M_{\text{sub}}$ is considered to be the subhalo mass at infall (relevant for \shmr) or the subhalo mass at a specific epoch. By assuming that the mass--to--light ratio of the satellites is linearly proportional $M_* \propto L$; and the $M_* - M_{\text{halo}}$ relation follows the power law $M_* \propto M_{\text{halo}}^{\beta}$, which is similar for satellites and hosts, we can write the subhalo mass function as
\begin{align}
    \frac{dN_{\text{sub}}}{dL} &\propto \frac{dN_{\text{sub}}}{dM_{\text{sub}}} \frac{dM_{\text{sub}}}{dL} \\
    &\propto L^{\frac{\alpha}{\beta}} L^{\frac{1-\beta}{\beta}}.
\end{align}
Assuming every subhalo is occupied by a satellite, we can compare the number of subhalos with our CLF using the luminosity form of the Schechter function $dN_{\text{sat}}/dL \propto L^{\alpha_s}$. This comparison allows us to match exponents and determine the low-mass end slope of $M_*-M_{\text{halo}}$ through the following equation 
\begin{align}
    \beta = \frac{1+\alpha}{1+\alpha_s},
    \label{eq: Beta Equation}
\end{align}
which has also been identified in previous studies such as \citet{garrison2014elvis}.

A steep slope $\beta$ for the low-mass end of the $M_*-M_{\text{halo}}$
relation means that as one approaches smaller halo sizes, star formation efficiency decreases rapidly. This could be driven by supernova feedback, which affects a galaxy's star formation efficiency \citep{shankar2006new}.
When the $M_*-M_{\text{halo}}$ relation has a steep slope $\beta$, hosts with very different stellar masses live in halos of roughly the same mass.

Typical values of $\beta$ range between \mbox{$\sim 1 < \beta < 2.5$} \citep{moster2010constraints, garrison2014elvis, behroozi2013average, Wang, Sales2013}, yet steeper values have also been found, e.g. $\beta = 3.1$ by \citet{shankar2006new} and \citet{brook2014stellar}; implying that their measured satellite luminosity functions had shallow slopes. 

Using our simple power-law matching of Equation \ref{eq: Beta Equation} to constrain the low-mass end slope of the \shmr~relation, we expect the values of $\beta$ to be negative if the CLF slope \mbox{$\alpha_s > -1$}, positive when \mbox{$\alpha_s < -1$}, and diverge as $\alpha_s$ approaches -1. 
Our central values of $\alpha_s$ are always $< -1$, as seen in Figure~\ref{fig:alpha_s}, and the probability of us inferring a value $< -1$ is 62\%, 82\%, and 71\% for the low-mass, high-$z$; high-mass, low-$z$; and high-mass, high-$z$ hosts, respectively.
Using the $\alpha_s$ parameter values that are $< -1$, we are able to set a lower limit on the low-mass end of the \shmr~slope to be \mbox{$\beta > 1.3$}; which is consistent with theoretical expectations, albeit with large uncertainty. 
If we were to use the values of $\alpha_s$ that are $> -1$, we would obtain a negative slope for the ~\shmr, implying that galaxies with large stellar masses occupy smaller halos. 
Although this result is not one we would expect, due to our inferred values of $\alpha_s$ we can not rule it out.
We have therefore shown that by measuring the shape of satellite luminosity function, one can independently set constraints on the slope of the \shmr~relation.

\section{Summary}
\label{sec:Summary}

We use a Bayesian statistical method, developed by \citet{Nierenberg2011}, to measure the satellite luminosity function of faint satellites with magnitudes \mbox{$m_{f814W} < 25$} around low-mass host galaxies at redshifts \mbox{$0.1 < z < 0.8$} for our high-mass sample \mbox{$10.0 < \log(M_*/M_{\odot}) < 10.5$}, and at redshifts \mbox{$0.4 < z < 0.8$} for our low-mass sample \mbox{$9.5 < \log(M_*/M_{\odot}) < 10.0$} and high-mass sample from the COSMOS survey. We examine trends in amplitude, redshift evolution, and shape of our cumulative luminosity functions, as a function of redshift, to those of \citet{Nierenberg2012} and \citet{Sales2013} for hosts of similar stellar masses. Our main results are summarized bellow.
\begin{enumerate}
    \item We can reliably measure the satellite luminosity function down to \mbox{$\Delta m = m_{\text{sat}} - m_{\text{host}} = 5.5$} for low-mass host galaxies at redshifts $0.4 < z < 0.8$; and $\Delta m = 3.5$ and $\Delta m = 4.0$ for high-mass host galaxies at low $(0.1 < z < 0.4)$ and high $(0.4 < z < 0.8)$ redshift respectively. These measurements are approximately equivalent to observing
    satellites with Fornax-like magnitudes for LMC-like luminosity hosts, out to redshifts of $0.8$.
    \item The amplitudes of our cumulative luminosity functions show the curves grouping together for the lower-mass hosts, especially at redshifts \mbox{$0.4 < z < 0.8$}, indicating that satellite abundance is independent of host stellar mass for low-mass hosts galaxies. 
    \item We do not detect any significant change in the CLF as a function of redshift within the measured uncertainties, in agreement with previous predictions.
    \item Using the slope of the CLF, we are able to constrain the low-mass slope of the $M_*$--$M_{\text{halo}}$ relation to $\beta > 1$.
\end{enumerate}
In the future, we can apply this approach to the much wider fields of view of the Roman Space Telescope High Latitude Survey \citep[HLS;][]{spergel2015} and the Legacy Survey of Space and Time \citep[LSST;][]{ivezic2019lsst}.
With a wider survey area, we will observe a far greater number of low-mass host galaxies and their satellites. A large number of high resolution images will also become available from the deep imaging of the Euclid Space Telescope \citep[Euclid;][]{racca2016euclid}, allowing us to push the limits to study even fainter galaxy systems.
With larger galaxy samples, we will need to rely on future spectroscopic facilities such as the Maunakea Spectroscopic Explorer \citep{mse2019}, for spectroscopic follow-up of these hosts. Deeper surveys will also allow us to detect even fainter satellites.
These future surveys play a crucial part in robustly measuring the CLF of isolated host galaxies, and using them to constrain the low-mass-end slope of the $M_* - M_{\text{halo}}$ function.

\section{Data Availability}
\label{sec:Data Availability}

The data and images used in this work were obtained from the COSMOS evolution survey, maintained by NASA/IPAC Infrared Science Archive, at
\url{https://irsa.ipac.caltech.edu/cgi-bin/Gator/nph-scan?mission=irsa&submit=Select&projshort=COSMOS}. All redshift values used can be found on the z-COSMOS @ LAM database available at \url{http://cesam.lam.fr/zCosmos/}.

\section{Acknowledgments}
\label{sec:Acknowledgments}

We thank Benjamin Buckman, Kirsten Casey, Bianca Davis, Jose Flores, Christopher Garling, Humberto Gilmer, Jahmour Givans, Johnny Greco, Laura Sales, and Franz Utermohlen for helpful comments. DR received support from the APS Bridge Program Fellowship.  AMN acknowledges support from the University of California Irvine, Chancellor's Postdoctoral Fellowship, as well as from the NASA Postdoctoral Program Fellowship, and from the Center for Cosmology and AstroParticle Physics Postdoctoral Fellowship.
AHGP is supported by National Science Foundation Grant Numbers AST-1615838 and AST-1813628. 
This work is based on zCOSMOS observations carried out using the Very Large Telescope at the ESO Paranal	Observatory	under Programme	ID:	LP175.A-0839, and 
on data products from observations made with ESO Telescopes at the La Silla Paranal Observatory under ESO programme ID 179.A-2005 and on data products produced by TERAPIX and the Cambridge Astronomy Survey Unit on behalf of the UltraVISTA consortium.



\bibliographystyle{mnras}
\bibliography{references}



\appendix


\bsp	
\label{lastpage}
\end{document}